\documentclass{aa}  
\usepackage{rotating}
\usepackage{graphicx}
\usepackage{txfonts}
\usepackage[colorlinks=true, citecolor = blue]{hyperref}

\begin{document} 
\definecolor{shockingpink}{rgb}{0.75, 0.06, 0.75}

   \title{Stability of Polycyclic Aromatic Hydrocarbon Clusters in Protoplanetary Disks}

   \subtitle{}

   \author{K. Lange
          \inst{1},
          C. Dominik
          \inst{1}
          \and
          A. G. G. M. Tielens
          \inst{2}$^\text{,}$\inst{3}}

   \institute{Anton Pannekoek Institute for Astronomy, University of Amsterdam,
              Science-Park 904, 1098 XH Amsterdam, Netherlands;
              \email{k.lange@uva.nl}
         \and
             {Leiden Observatory, Leiden University, P.O. Box 9513, 2300 RA Leiden, Netherlands}
        \and {Astronomy Department, University of Maryland, MD 20742, USA}
             }

   \date{Received xxx; accepted xxx}
 
  \abstract
   {The infrared signature of polycyclic aromatic hydrocarbons (PAHs) are present in many protostellar disks and these species are thought to play an important role in heating of the gas in the photosphere.}
   {We aim to consider PAH cluster formation as one possible cause for non-detections of PAH features in protoplanetary discs. We test the necessary conditions for cluster formation and cluster dissociation by stellar optical and FUV photons in protoplanetary disks using a Herbig Ae/Be and a T\,Tauri star disk model.}
   {We perform Monte-Carlo (MC) and statistical calculations to determine dissociation rates for coronene, circumcoronene and circumcoronene clusters with sizes between 2 and 200 cluster members. By applying general disk models to our Herbig Ae/Be and T\,Tauri star model, we estimate the formation rate of PAH dimers and compare these with the dissociation rates.}
   {We show that the formation of PAH dimers can take place in the inner 100\,AU of protoplanetary disks in sub-photospheric layers. Dimer formation takes seconds to years allowing them to grow beyond dimer size in a short time. We further demonstrate that PAH cluster increase their stability while they grow if they are located beyond a critical distance that depends on stellar properties and PAH species. The comparison with the local vertical mixing time scale allows a determination of the minimum cluster size necessary for survival of PAH clusters.}
   {Considering the PAH cluster formation sites, cluster survival in the photosphere of the inner disk of Herbig stars is unlikely because of the high UV radiation. For the T\,Tauri stars, survival of coronene, circumcoronene and circumcircumcoronene clusters is possible and cluster formation should be considered as one possible explanation for low PAH detection rates in T\,Tauri star disks.}

 
   \titlerunning{Stability of Polycyclic Aromatic Hydrocarbon Clusters}
   \authorrunning{Lange et al.}
   \maketitle
   
\section{Introduction}
Prominent infrared (IR) features of polycyclic aromatic hydrocarbons (PAHs) \citep{Allamandola1989} have been widely observed in the interstellar-medium (ISM) (see \citet{Tielens2008} for a review) and protoplanetary disks \citep{Acke2010, Geers2007}.
PAHs are strong absorbers in ultraviolet (UV) and are responsible for the heating of gas in the ambient medium \citep{Bakes1994} which is of interest for disk evaporation models \citep{Rodenkirch2020}.
\citet{Lagage2006} and \citet{Doucet2007} used PAH emission imaging to characterise the flaring of the protoplanetary disk HD\,97048.
Furthermore, the relative intensity of the $6.2$\,$\mu$m and $11.3$\,$\mu$m spectral features makes it possible to trace the ionisation state of PAHs and therefore probe their radiative environment \citep{Hony2001, Peeters2002}.
As PAH molecules are small, they are coupled to the gas and can be used to trace gas mixing processes. 
Therefore, the study of PAHs in protoplanetary disks can be one step towards determining the unknown turbulence parameter $\alpha$ \citep{Shakura1973} in protoplanetary disks.\\
\\From a statistical point of view, around 60\% of disks around Herbig Ae/Be stars \citep{Acke2004} but less than 10\% of disks around T\,Tauri stars \citep{Geers2006} show PAH features in their spectra. This issue has been discussed in \citet{Siebenmorgen2010} and \citet{Siebenmorgen2012} and was attributed to the higher UV intensity in Herbig stars and the destruction of PAH molecules by X-ray absorption in T\,Tauri stars. Vertical mixing has been identified as one mechanism for survival by mixing PAHs into optically thick layers before they are destroyed. \citet{Geers2009} argue, that the lack of PAH emission can be a result of freeze-out on dust grains.\\
\\Besides photodissociation, clustering can hamper PAH detection. Due to their larger size, clusters are cooler. As the emission strength of the aromatic infrared bands (AIB) is very sensitive to the temperature (e.g. \cite{Leger1984}), cluster have weaker emission in the short wavelength bands compared to individual molecules. Especially with the 10\,$\mu$m and 18\,$\mu$m silicate emission, this makes it harder to identify PAHs from the continuum in protoplanetary disks.\\
\\PAH clustering has been studied in the context of soot particle formation in combustion engines (e.g. see \citet{Raj2010} or \citet{Sander2011}) as well as in the context of astronomy. Molecular dynamic simulations show that the outcome of two colliding PAH molecules is likely a molecule cluster bound by van der Waals forces if the collision energy is low \citep{Rapacioli2006}. In order to reach 50\,\% bouncing probability for a monomer-monomer collision, the ambient gas needs a temperature of 13 500\,K and even higher temperatures are needed if larger structures are involved in a collision. \citet{Rapacioli2006} also report that cluster structures can dissociate under incident UV radiation and liberate individual molecules again.\\
\\Because of their potential as tracers of turbulent mixing, we want to study the evolution of PAHs in protoplanetary disks. Such an endeavour will have to include a deep understanding of the processes involved in the formation and destruction of PAH clusters. This will be the basis for a follow up study that will also include photo destruction of PAH monomers and their vertical and radial transport through the disk. Such a study will be very timely given the upcoming launch of the James Webb Space Telescope (JWST) as the Mid-Infrared Instrument (MIRI) is well suited to probe the mid-IR spectra of protoplanetary disks with unprecedented sensitivity and spatial resolution.\\
\\This work is organised as follows: in section \ref{sec:MC}, we introduce our PAH model to simulate cluster formation and their dissociation. In section \ref{sec:diskmodel} we explain our simplified disk model in which our PAHs are located in. In section \ref{sec:results}, we will present our main results of our comparison between dissociation and clustering and we will further investigate the stability of larger clusters in disk environments. In sections \ref{sec:discussion} and \ref{sec:conclusions} we discuss and summarise our results.

\section{Methods}
\label{sec:MC}
\subsection{Polycyclic Aromatic Hydrocarbons}
PAH molecules belong to the group of hydrocarbons that contain two or more aromatic rings with delocalised $\pi$ electrons. The family of PAHs is huge, covering many molecule sizes from the smallest molecule naphthalene C$_{10}$H$_8$ up to C$_{348}$H$_{48}$ with many possible isomers.\footnote{see e.g. the PAH IR spectral database for an overview of PAH molecules \url{https://www.astrochemistry.org/pahdb/}.} In general, the sharp PAH features have been attributed to smaller PAHs with typically 50 C atoms because their heat capacity is low enough to allow MIR emission \citep{Allamandola1989}. Since individual features are only correlated to specific bonds (e.g. $3.3$\,$\mu$m feature: CH stretching mode, $6.2$\,$\mu$m feature: CC stretching mode \citep{Tielens2008}, the exact molecular structure and size distribution of astronomical PAHs is unknown. Therefore, assumptions concerning size and structure are needed.
An exception to this is Buckminsterfullerene (C$_{60}$) with its unique spectral features located at 7.0, 8.5, 17.4 and 19.0$\,\mu$m \citep{Menendez2000} which have been identified in planetary nebulas \citep{Sellgren2010, Cami2010}.\\
\\For our model, we choose to simulate the three molecules coronene (C$_{24}$H$_{12}$), circumcoronene (C$_{54}$H$_{18}$) and circumcircumcoronene (C$_{96}$H$_{24}$) to account for the range of astrophysical relevant sizes \citep{Allamandola1989, Croiset2016}. Because of their compact circular structure, these molecules are more stable compared to other PAHs of similar size \citep{Tielens2008}. Hence, these molecules have a higher probability to survive under the conditions of the ISM and protoplanetary disks. In our model studies, we adopt, for simplicity, that each disk model only hosts one specific PAH species instead of assuming molecule size distributions. This allows us to assess general trends related to molecule size while keeping the model as simple as possible.\\
\\Furthermore, we define a PAH cluster as any molecular structure with more than one PAH monomer, bound by van der Waals forces. As calculated by \citet{Rapacioli2005}, the lowest energy structures of small clusters are stacks while larger cluster consist of structures made of several stacks. The number of energetically favoured configurations of the member molecules is temperature dependent and related to misalignment and reorganisation of individual molecules \citep{Rapacioli2007}. 
When evaluating the dissociation rate of PAH cluster, the rate at which monomers detach, the binding energy of the monomer to the cluster is decisive for lifetime of the cluster. Since it is impractical to follow the exact configuration of the monomers in all cluster sizes (single stacks, stack agglomerations, 3D stack structures) and therefore their exact binding energy, we have to assume an average binding energy as the sublimation energy of the corresponding molecule in equation \eqref{eqEA}.\\
\\The photochemistry of PAHs in an astrophysical environment is controlled by the internal energy of the molecule. Hence, we have to study its temperature behaviour which is a balance between absorbed stellar UV and visible photons and the PAH energy loss by IR photon emission and photochemical processes. In this work, we only consider IR photon emission and photodissociation of a PAH cluster as energy loss channels. To statistically derive microcanonical temperatures, we simulate individual heating and cooling events of a PAH cluster with a Monte-Carlo (MC) scheme. In this section, we will first describe all considered events individually and then explain the Monte-Carlo scheme.\\
\subsection{Monte-Carlo Analysis}
\subsubsection{Photon Absorption Rate}
The absorbed energy flux $\Phi_{\text{a},\text{E}}$ at photon energy $E$ is 
\begin{equation}
    \Phi_{\text{a},\text{E}} = F_\text{E} \sigma_\text{E}
\end{equation}
with $F_\text{E}$ being the flux density and $\sigma_\text{E}$ being the absorption cross section of PAHs at photon energy $E$. We calculate $F_\text{E}$ from a black body spectrum with stellar parameters given in table \ref{table:stars} and use $\sigma_\text{E}$ from time dependent density functional theory (DFT) calculations performed by \citet{Malloci2007}\footnote{available at \url{https://www.dsf.unica.it/~gmalloci/pahs/pahs.html}}. Note that no spectrum is available for circumcircumcoronene. 
Since the absorption spectrum is size and PAH family dependent, we use the circumcoronene spectrum for circumcircumcoronene instead of extrapolating the spectrum from circumcoronene or choosing the largest available molecule that does not belong to the coronene family. Additionally, for our T\,Tauri model we account for the FUV excess due to accretion by adding 1\% of the total energy flux \citep{Siebenmorgen2010} as Lyman-alpha emission at 10.2\,eV. We divide the energy space into $i$ bins with average energy $E_i$. The rate of absorbed photons of a monomer in wavelength interval $i$ is then
\begin{equation}
    r_{\text{a}, i} = \frac{1}{E_i} \int_{E_\text{i, min}}^{E_\text{i,max}} \Phi_{\text{a}, \text{E}} \text{d}E \text{.}
\end{equation}
The photon absorption rate of clusters can be determined by scaling the absorption rate of a monomer to the number of monomer molecules in the PAH cluster.
We choose 100 energy bins\footnote{We do not bin in wavelength space to ensure a linear step size in the energy grid.} between 0.1\,eV and 13.6\,eV. Our upper energy is determined by the hydrogen ionisation energy above which the optical depth increases by orders of magnitude \citep{Ryter1996}. These photons cannot penetrate the disk deeply and therefore do not contribute to the energy budget of PAHs. 

\subsubsection{Photon Emission Rate}
To model the photon emission, we use the general expressions for cooling and microcanonical temperature $T_m$ given in \citet{Bakes2001} and \citet{Tielens2020} and extend these to lower temperatures. As stated before, we want to focus on large trends rather than determining exact numbers. The adopted cooling law is given by
\begin{equation}
    \frac{\text{d}T_m}{\text{d}t} = -1.1 \cdot 10^{-5}T_m^{2.53} \text{\,K/s}
    \label{eq:dTdt}
\end{equation}
and we relate that to the internal energy $E$ using\footnote{In order to preserve differentiability at the transition we use the smoothening relation $T_\text{m}=(T_\text{m,1}^8+T_\text{m,2}^8)^{0.125}$. We have tested both shallower and steeper power laws and find that this relation is the best compromise between smoothness and deviation from the original curves around the transition point.}
\begin{equation}
    T_\text{m} = \begin{cases}
    3750 \left(\frac{E\text{(eV)}}{3N-6}\right)^{0.45}\text{\,K}\text{\hspace{1cm} if $T_m < 1000$\,K}\\
    11000 \left(\frac{E\text{(eV)}}{3N-6}\right)^{0.8}\text{\,K}\text{\hspace{1cm} if $1000\text{\,K} < T_m$}\\
    \end{cases}
    \label{eq:T(E)}
\end{equation}
Since we apply this relation to monomers and clusters, $N$ describes the total number of atoms in the given monomer or cluster. We use a fixed photon energy of $E_\text{IR}=0.145$\,eV which is equal to the energy to photons of the 8.6\,$\mu$m feature. Finally, we compute the average photon emission rate $r_\text{e}$ by combining equations \eqref{eq:dTdt} and \eqref{eq:T(E)} to obtain d$E$/d$t$ and using
\begin{equation}
    r_\text{e} = \frac{\text{d}E}{\text{d}t} E_\text{IR}^{-1} \text{.}
\end{equation}

\subsubsection{Dissociation Rates}
\label{sec:N=N+1}
\begin{figure}
    \centering
    \includegraphics[width=1\linewidth]{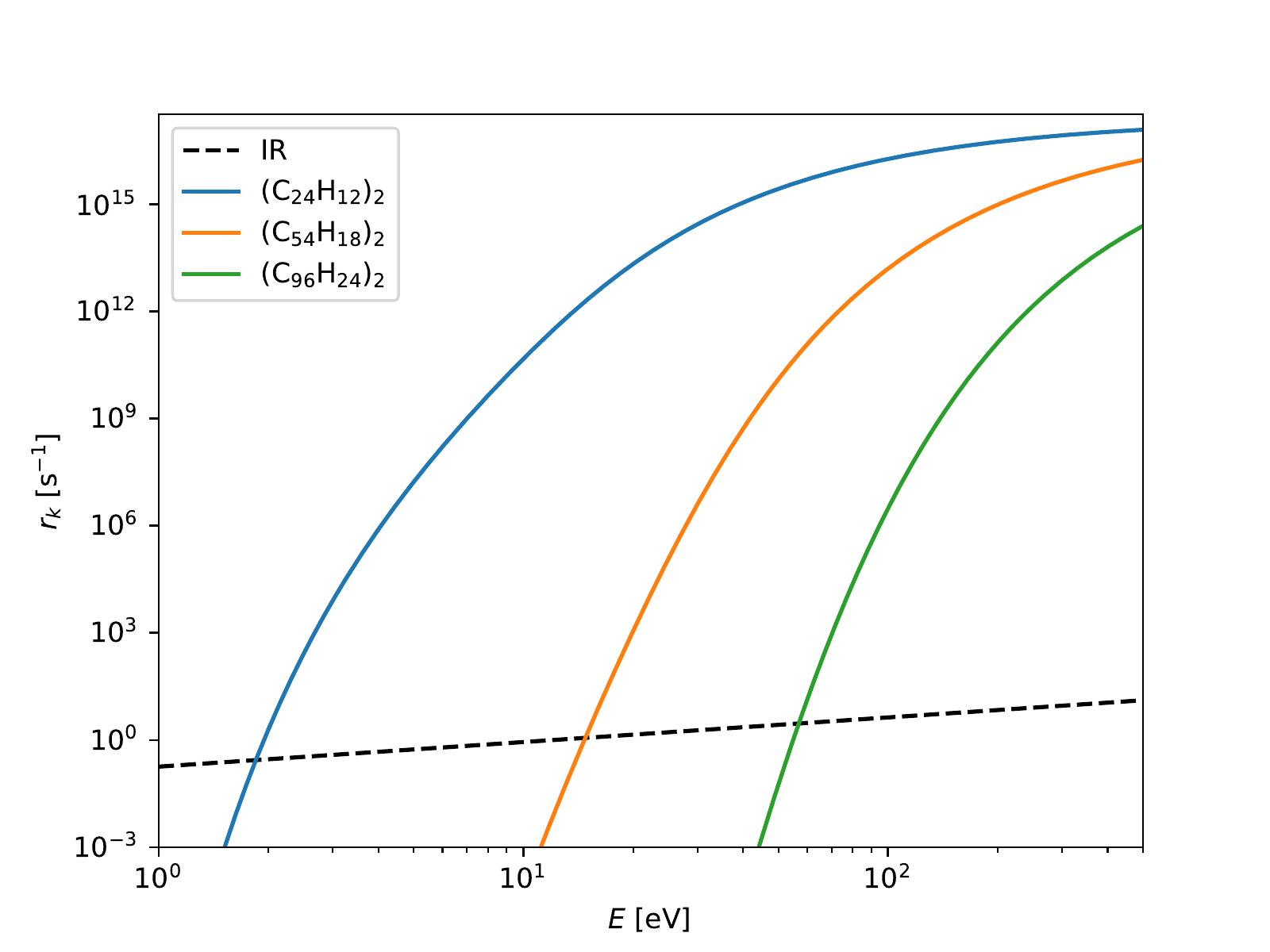}
    \caption{Comparison between photodissociation rates obtained from equation \eqref{eq:arrhenius} of PAH dimers and IR cooling rates ,  \textit{dashed}) for the three simulated molecules coronene (\textit{blue}), circumcoronene (\textit{orange}) and circumcircumcoronene (\textit{green}). The transition from IR cooling as the dominant energy loss channel to photodissociation is sharp.}
    \label{fig:k(E)}
\end{figure}
To calculate the dissociation rates of a PAH cluster for single and multi-photon processes, the individual rates are evaluated as a function of energy from an Arrhenius expression derived from Rice-Ramsperger-Kassel-Marcus theory (RRKM theory). We follow the approach of \citet{Tielens2005} and use
\begin{equation}
    r_\text{k} = k_0\cdot \text{exp}\left(\frac{-E_\text{A}}{k_\text{B}T_\text{e}}\right)
    \label{eq:arrhenius}
\end{equation}
with the entropy factor $k_0$, the activation energy $E_\text{A}$, Boltzmann's constant $k_B$ and the effective temperature $T_\text{e}$. $T_\text{e}$ can be calculated from the microcanonical temperature by adding a heat bath correction derived from the density of states and becomes 
\begin{equation}
    T_\text{e} = T_\text{m} \left(1-0.2\frac{E_\text{A}}{E}\right) \text{\,.}
\end{equation}
For $k_0$, we assume a value of $k_0 = 2.5\cdot10^{17}$\,s$^{-1}$ for all three PAH monomers which is in agreement with the values given in \citet{Zacharia2004}. A derivation of $k_0$ can be found in the appendix. Instead of using the graphite exfoliation energy \citep{Herdman2008} we calculate the activation energy $E_A$ for PAH cluster dissociation by using an approximation from DFT calculations for the adsorption of PAHs onto graphene by \citet{Li2018} 
\begin{equation}
    \frac{E_\text{A}}{N_\text{C}} = 0.046+0.021\frac{N_\text{H}}{N_\text{C}}
\label{eqEA}
\end{equation}
which is close to the approximation obtained from the experimental study of \citet{Zacharia2004}.
For our three considered molecules this corresponds to $E_\text{A}=1.4$\,eV for C$_{24}$H$_{12}$, $E_\text{A}=2.9$\,eV for C$_{54}$H$_{18}$ and $E_\text{A}=4.9$\,eV for C$_{96}$H$_{24}$.  
The resulting dissociation rates for C$_{24}$H$_{12}$, C$_{54}$H$_{18}$ and C$_{96}$H$_{24}$ dimers are displayed in figure \ref{fig:k(E)} together with the IR photon emission rate.\\
\\We account for the energy loss during dissociation by distributing the remaining energy of the cluster $E' = E - E_\text{A}$ after break-up of one molecule equally to the degrees of freedom of the remaining daughter cluster and the separated monomer following:
\begin{equation}
    E_\text{dau} = \frac{3N_\text{dau}-6}{3(N_\text{dau}+N_\text{mon})-6} E' \text{\hspace{0.5cm}and\hspace{0.5cm}} E_\text{mon} = E' - E_\text{dau}
    \label{eq:energy_fraction}
\end{equation}
where $N_\text{dau}$ is the number of atoms in the daughter cluster and $N_\text{mon}$ is the number of atoms in the monomer.\\
\\After one dissociation event, by definition, the cluster with $N_\text{mem}$ member molecules has changed its size and does not belong to the same sampled species anymore. Strictly speaking, the energy probability distribution after dissociation needs to be taken into account for calculating the energy probability distribution of the species with one molecule less. This would require recursive calculations starting with the largest cluster while taking into account the local history of PAHs (multiple growth channels, dissociation). To properly sample it, a full cluster size evolution model is required. We have elected to ignore this change in the size of a dissociating PAH and assume that larger cluster with $N_\text{mem}+1$ monomers that fragment into the sampled cluster size with $N_\text{mem}$ members have the same energy as the remaining energy of the daughter cluster with $N_\text{mem}-1$ monomers after fragmentation from a cluster with $N_\text{mem}$ monomers.

\subsubsection{Monte-Carlo Scheme}
Our Monte-Carlo scheme follows the fundamental principles explained in \citet{Zsom2008}. Given the calculated rates described above, the total rate of events for a simulated PAH molecule is
\begin{equation}
    r = \sum_i r_{\text{a}, i} + r_\text{e} + r_\text{k} \text{\,.}
\end{equation}
We determine a MC time step $\delta t$ by drawing it randomly from an exponential distribution with mean $1/r$
\begin{equation}
    f(\delta t) = r \,\text{exp}(-r\delta t) \text{.}
\end{equation}
To choose which event has occurred we define the set of all possible events with rates $R = \{ r_\text{{a},i}$, $r_\text{e}, r_\text{k}\}$. The probability of event $j$ with rate $r_j \in R$ occurring is then
\begin{equation}
    P(j) = \frac{r_j}{r} \text{.}
\end{equation}
Finally, the average dissociation rate of the simulated PAH cluster can be obtained from integration of the dissociation rate over the energy probability distribution\footnote{Note that the dissociation rate can be also obtained from counting the dissociation events but the integral is more stable towards MC noise}
\begin{equation}
    k = \int_0^{\infty} G(E)r_\text{k}(E) \text{d}E
    \label{eq:k_int}
\end{equation}
following \citet{Tielens2005}. We run our MC model in blocks of $10^5$ events and determine the energy probability distribution $G(E)$ and the dissociation rate $k$ after each block. If the standard deviation of the mean value for $k$ is less than 5\,\%, we stop our calculations. If not, we sample until we have simulated $2\cdot10^{7}$ events. In these cases, the dissociation rate is very low and dominated by rare high-energy events and the precise value is not astronomically relevant.

\begin{figure}
    \centering
    \includegraphics[width=1\linewidth]{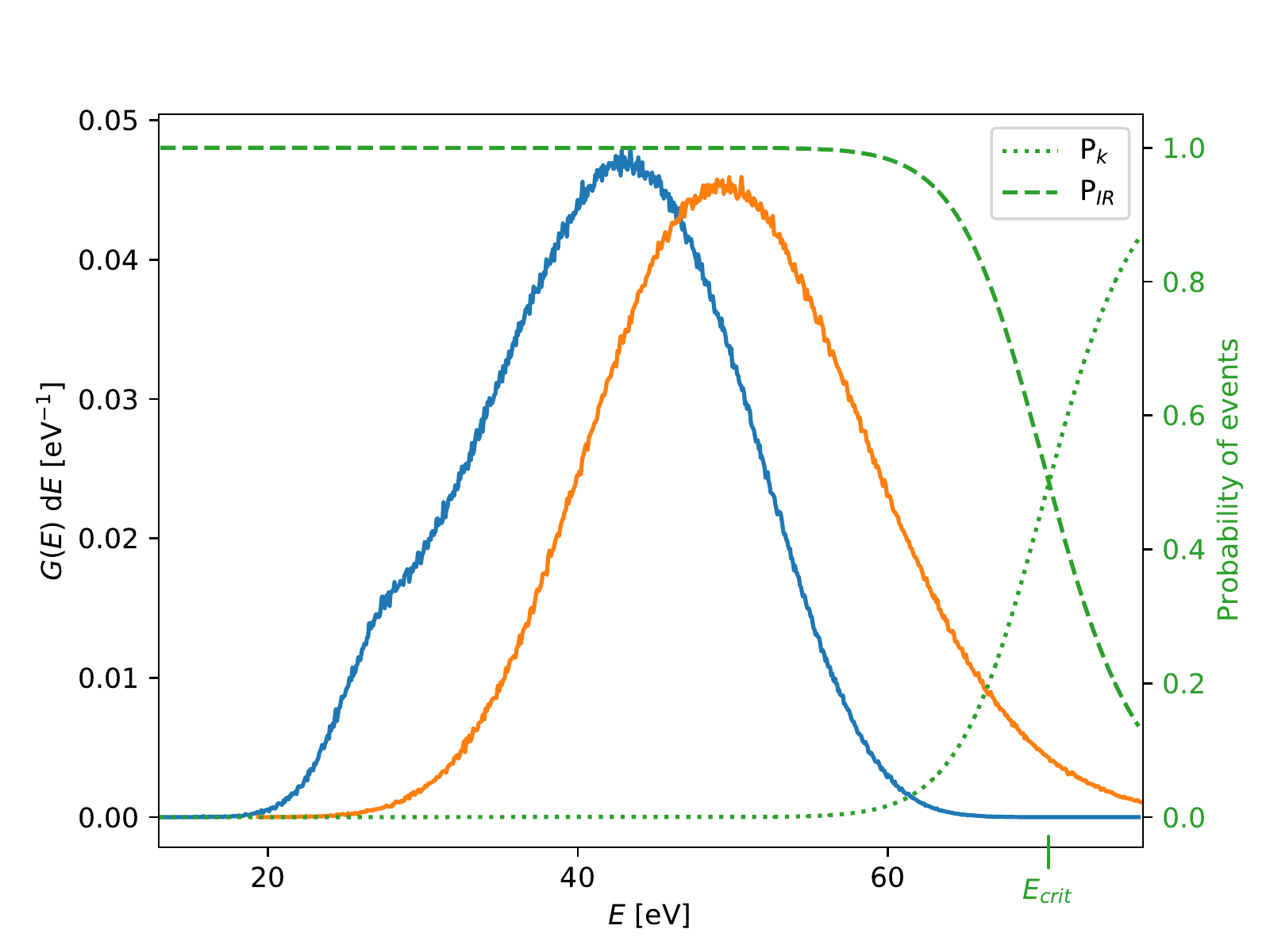}
    \caption{Comparison of the energy probability distribution $G(E)$ obtained from Monte-Carlo (Herbig star model, 7.5\,AU, (C$_{96}$H$_{24}$)$_2$) with photodissociation (\textit{blue}) and without photodissociation (\textit{orange}). Additionally, the probability for photodissociation (\textit{dotted}) and IR-cooling (\textit{dashed}) are shown. With photodissociation, PAHs cool down so efficiently that only a small fraction of time is spent at high temperatures.}
    \label{fig:MC_GE}
\end{figure}

\subsection{Statistical Analysis}
To confirm our MC results where the dissociation rate is very small, we perform a statistical analysis based on the works of e.g. \citet{Purcell1976}, \citet{Bakes2001} and \citet{Andrews2016}. Since cluster dissociation events are rare in these cases, they will not contribute significantly to the energy probability density. Hence, we neglect energy loss due to dissociation in the statistical analysis.\\
\\Assuming that absorbed and emitted photons are Poisson distributed, the internal energy probability distribution $G(E)$ can be described by 
\begin{equation}
    G(E)\text{d}E = \frac{\bar{r}}{\text{d}E/\text{d}t} \text{exp}\left(-\bar{r}\tau_{\text{min}}(E) \right)\text{d}E
    \label{eq:G(E)eq}
\end{equation}
where $\bar{r}=\sum_i r_{\text{a},i}$ is the average photon absorption rate, $\frac{\text{d}E}{\text{d}t}$ is derived from eq.\eqref{eq:dTdt} and \eqref{eq:T(E)} and $\tau_{\text{min}}$ is the minimum time a PAH needs to cool from an excitation energy $E_1$ to $E$. We consider the photon energy distribution in the same way as in the MC method by binning the spectrum. $\Tilde{G}(E)$ is then obtained by averaging $G(E)$ over all photon energies $E_i$ by
\begin{equation}
    \Tilde{G}(E) = \frac{1}{\bar{r}} \sum_i G(E, E_i) \, r_{\text{a},i}.
\end{equation}
Because we focus on multi-photon events in this analysis, the energy probability distribution $G_n(E)$ has to be iteratively calculated with 
\begin{equation}
    G_n(E) = \int_0^\infty G_{n-1}(E_\text{eq},\sigma_E,E')\Tilde{G}(E,E')\text{d}E' \text{\,.}
\end{equation}
 We set our initial energy probability distribution $G_1$ to a Gaussian distribution around the radiative equilibrium energy $E_\text{eq}$ with standard deviation $\sigma_E$ to increase the convergence speed (see the appendix for an analytical approach and the derivation of $E_\text{eq}$ and $\sigma_E$). Once convergence is achieved, we calculate the cluster dissociation rate according to eq. \eqref{eq:k_int}.\\
\\Figure \ref{fig:MC_GE} shows the energy probability distribution for a C$_{96}$H$_{24}$ dimer obtained from the MC approach with cluster dissociation (\textit{blue}) and without (\textit{orange}). Dissociation leads to efficient cooling and only a small fraction of time (high-energy tail) is spent at energies where cluster dissociation is dominant. Additionally, the critical energy $E_\text{crit}$, where the probability for cooling and dissociation are equal, is indicated by the green probability curves. Internal energies between 50\,eV and 60\,eV become rare because these are results of cooling from higher energies. Since dissociation is dominant at high energies, the dimer does not continuously cool down but instead loses half of its energy due to dissociation and is the origin of the shoulder at 30\,eV.\\

\section{The disk model}
In order to investigate the importance of clustering and cluster dissociation in protoplanetary disks, we set-up two specific disk models. For our Herbig model, we have chosen HD169142 and for the T\,Tauri model BP Tau for direct comparison. We do not aim to model the observed disks in both system with their features but focus on taking the host stars as representative stars for Herbig resp. T\,Tauri stars to study the effects of the radiative environment. We therefore apply a very general disk model. The stellar parameters for both stars are given in table \ref{table:stars}.
\label{sec:diskmodel}


\subsection{Disk profiles}

\begin{table}
\caption{Stellar parameters used for the disk models.$^1$: \citet{Honda2012}; $^2$:\citet{Grankin2016}}.
\begin{tabular}{l|c|c|c|c}
    Model & Reference star & $M_*$ & $L_*$ & $T_\text{eff}$\\
    \hline
     Herbig & HD169142$^1$ & 2.28 M$_\odot$& $15.33\,$L$_\odot$ &8200\,K\\
     T\,Tauri & BP Tau$^2$ & 0.75 M$_\odot$& $0.88\,$L$_\odot$& 4040\,K\\
\end{tabular}
\label{table:stars}
\end{table}

We apply our dissociation rate calculations to a protoplanetary disk with a surface density profile 
\begin{equation}
    \Sigma(r) = 730 \frac{\text{g}}{\text{cm$^2$}} \left(\frac{r}{1\,\text{AU}}\right)^{-1.5} \left(\frac{M_*}{M_\odot}\right)
\end{equation}
following a Minimum Mass Solar Nebula (MMSN) powerlaw profile as prescribed by \citet{Weidenschilling1977}. Our disk is set-up that the inner 100\,AU contain a total mass of 0.01$M_*$. We assume a vertical gas density profile in hydrostatic equilibrium 
\begin{equation}
    \rho(z,r) = \rho_\text{mid}(r)\,\text{exp}\left(\frac{-z^2}{2H^2}\right)
\end{equation}
where $H$ denotes the gas pressure scale height given by $H=c_\text{s}/\Omega_\text{K}$ and $\rho_\text{mid}$ is calculated with $\rho_\text{mid}(r) = \Sigma(r)/\sqrt{2\pi}H(r)$. The isothermal sound speed is given by the expression $c_s=\sqrt{k_\text{B}T/\mu m_\text{p}}$ where we assume $\mu = 1.37$.
The disk follows an isothermal temperature profile in vertical direction
\begin{equation}
    T(r) = 280\,\text{K} \left(\frac{r}{\text{AU}}\right)^{-0.5}\left(\frac{L_*}{L_\odot}\right)^{0.25}
\end{equation}
formulated by \citet{Hayashi1981}.\\
\\We furthermore assume that the dust is well mixed with the gas and has a composition similar to that of the ISM to estimate the extinction caused by dust and gas in the disk. Our estimated value for the UV extinction cross section normalised to H-atoms is $\sigma_\text{UV} \approx 10^{-21}$\,cm$^2$ per H atom \citep{Ryter1996}. We determine the optical depth by integrating the hydrogen density along the line of sight. The extinction cross section depends on the grain size distribution and local dust to gas ratio. To estimate the effects of dust settling, we vary $\sigma_\text{UV}$ by one order of magnitude in both directions to demonstrate the shifting of the $\tau_\text{UV}$=1 surface in the disk.

\begin{figure*}
    \centering
    \includegraphics[width=1\linewidth]{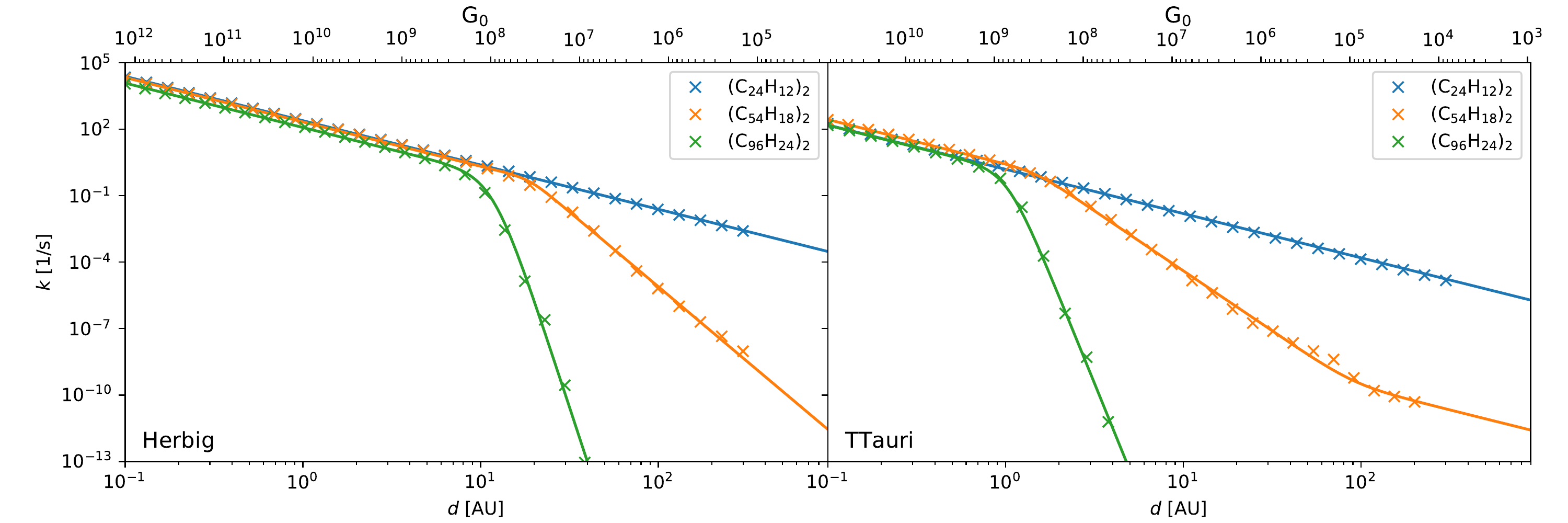}
    \caption{Dimer dissociation rates obtained by Monte-Carlo simulations (\textit{crosses}) for coronene (\textit{blue}), circumcoronene (\textit{orange}) and circumcircumcoronene (\textit{green}) calulated with eq. \ref{eq:k_int} with stellar parameters given in table \ref{table:stars}. In both cases, coronene dimers can be dissociated with single photon events. For larger molecules, the dissociation rate decreases tremendously with $G_0$ when the radiative equilibrium energy drops below the critical energy. The multi power-law fits (\textit{lines}) are used for interpolation in figure \ref{fig:k/Zn}.}
    \label{fig:k(d)}
\end{figure*}

     \begin{figure*}
        \includegraphics[width=1\hsize]{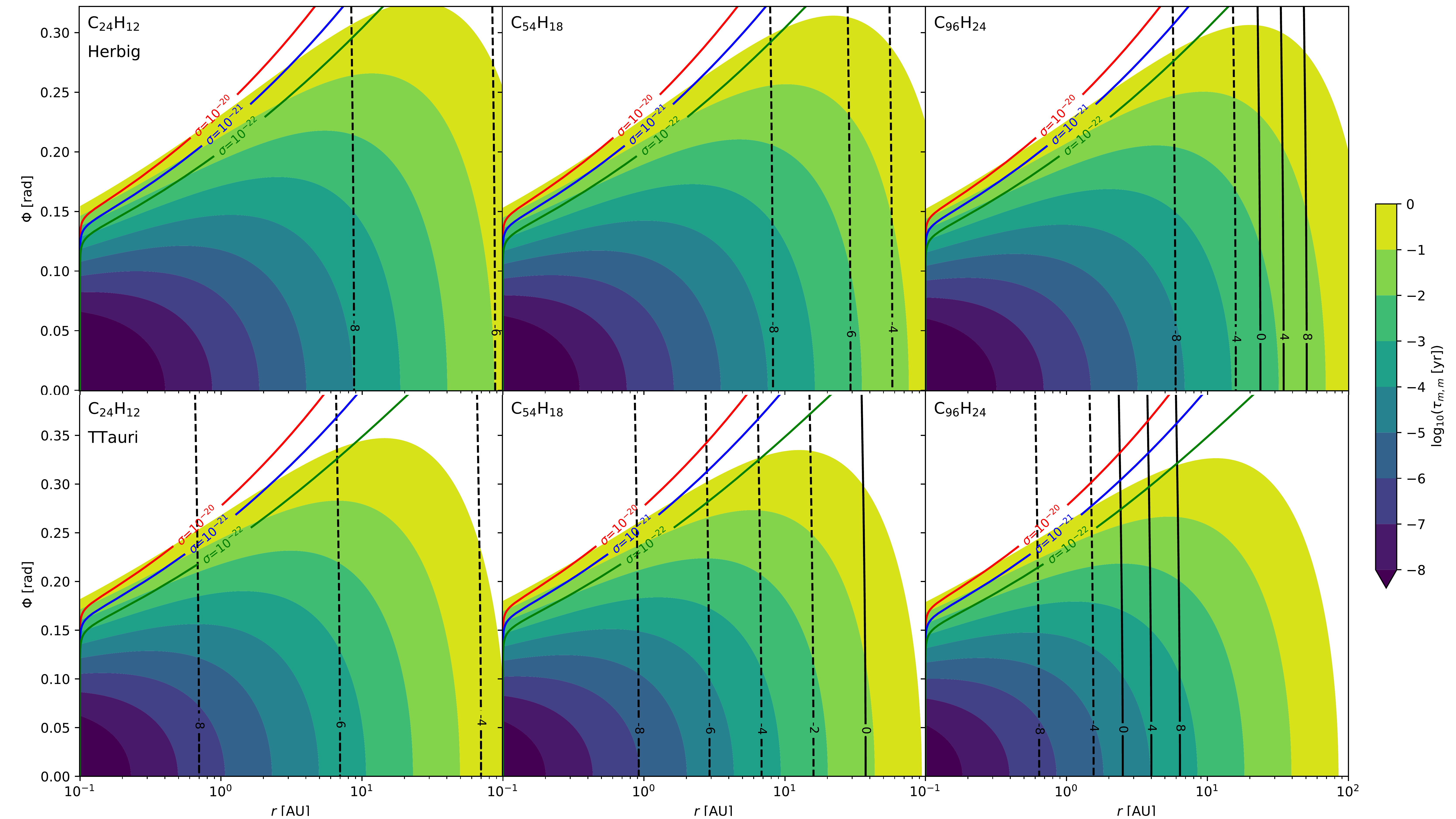}
            \caption{Comparison between collision time scale of monomers (\textit{filled contours}) and dissociation time scale of dimers (\textit{black contours}, same units) for the Herbig model (\textit{top row}) and the T\,Tauri model (\textit{bottom row}). \textit{Left}: coronene, \textit{middle}: circumcoronene, \textit{right}: circumcircumcoronene. Estimated $\tau_\text{UV}=1$ surfaces are indicated by coloured lines. Monomer collisions occur in all cases in disk relevant time scales within 100\,AU below the $\tau = 1$ surfaces.}
         \label{fig:k/Zn}
   \end{figure*}

\subsection{PAH monomer collisions}
For our model we assume that our disk gas and dust composition is similar to that of the interstellar medium (ISM). We therefore use the proposed carbon-atoms-locked-in-PAHs to hydrogen atom ratio that is obtained from the ISM $\text{C}_\text{PAH}:\text{H} = 1.5\cdot10^{-5}$ given in \citet{Tielens2008} to calculate number densities $n_\text{PAH} = n \cdot 1.5\cdot10^{-5} / N_\text{C}$ of our simulated PAH molecules with $N_\text{C}$ C atoms from the hydrogen density $n$ within the disk. We neglect PAH formation in the disk and assume that all PAHs are primordial. Thermal motion is the only cause of collisions between two PAHs in the gas phase. The relative velocities between gas-phase PAH molecules can then be expressed by 
\begin{equation}
    \Delta v = \sqrt{\frac{8k_\text{B}T}{\pi\mu_\text{PAH}}}
\end{equation}
with $\mu_\text{PAH}$ being the reduced mass of two colliding PAH molecules and $T$ being the kinetic temperature of the gas. Since our focus lies on collisions between individual PAH molecules we assume that PAH monomers are planar and their molecule radius can be expressed with $r_\text{PAH}=0.9 \sqrt{N_\text{C}}\,\AA$ where $N_\text{C}$ is the number of carbon atoms in a monomer \citep{Tielens2008}. Hence, the maximum geometrical collision cross section assuming parallel orientation for monomer-monomer interaction is
\begin{equation}
    \sigma_\text{m,m} = (1.8 \AA)^2 \pi  N_\text{C} \text{.}
\end{equation}
Finally, combining all expressions, the collision rate per PAH molecule is given by
\begin{equation}
    k_\text{PAH} = n_\text{PAH} \sigma_\text{m,m} \Delta v 
    \label{eq_Z/n}
\end{equation}
and we define the collision time scale per PAH molecule as 
\begin{equation}
    \tau_{m,m} = k_\text{PAH}^{-1} \text{.}
\end{equation}
Note that all quantities in equation \eqref{eq_Z/n} depend on $N_\text{C}$ and are therefore PAH molecule specific.


\begin{figure}
    \centering
    \includegraphics[width=1\linewidth]{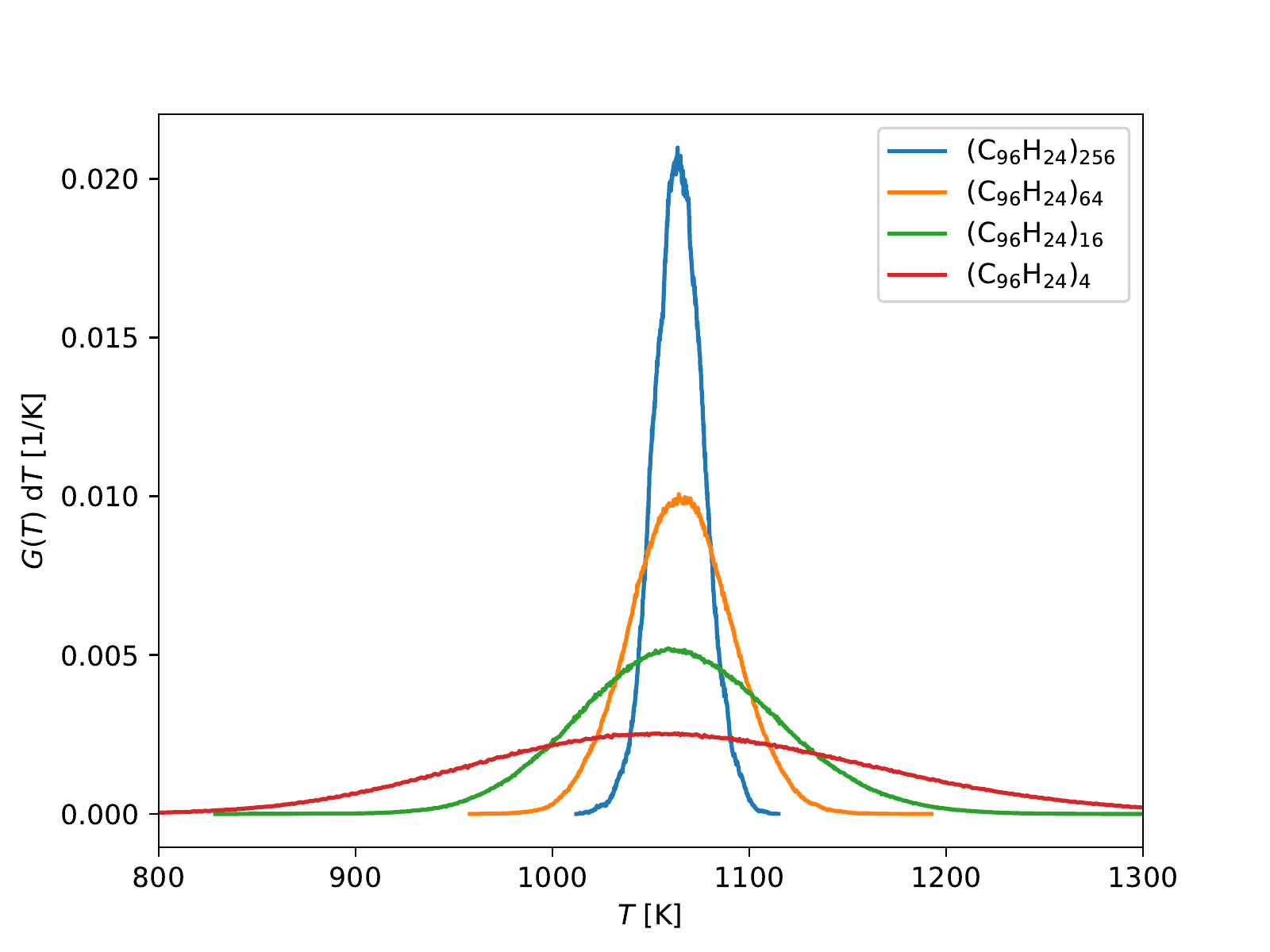}
    \caption{Temperature probability distribution in the same environment ($d=12.5$\,AU in HD169142) for different circumcircumcoronene clusters. With growing size, temperature fluctuations decrease because of a higher photon absorption rate and the dissociation rate converges towards the value at the radiative equilibrium temperature.}
    \label{fig:GE_gauss}
\end{figure}

\section{Results}
\label{sec:results}
\subsection{Formation and destruction of dimers}
We want to compare the PAH dimer dissociation time scale to the clustering time scale of monomers in our two disk models. 
Therefore, we ran our Monte-Carlo scheme for several radiation field intensities $G_0$ (in units of the mean Habing field) and computed the dissociation rates for coronene, circumcoronene and circumcircumcoronene.
Figure \ref{fig:k(d)} shows the resulting dimer dissociation rates as a function of $G_0$ and the corresponding distance from the star in the disks.
The heat capacity of coronene dimers is small enough so that single photons heat the dimer above the critical energy $E_\text{crit}$ and trigger dissociation. Hence, the dissociation rate decreases proportional to $G_0$ resp. the squared distance.
In contrast, circumcoronene and circumcircumcoronene dimers require multi-photon events and cannot be dissociated by individual photons. Close to the star, the radiative equilibrium energy is higher than $E_\text{crit}$ and the internal energy will strive through many photon absorptions towards $E_\text{crit}$ and finally dissociate the cluster. Consequently, the dissociation rate decreases linear with $G_0$ until $E_\text{eq}$ drops below $E_\text{crit}$ in weaker radiation fields. Then, the high-energy tail of $G(E)$ dominates the dissociation rate leading to a strong decrease given the exponential shape of the Arrhenius law. Additionally, the Lyman-alpha photons in the T\,Tauri disk are able to trigger single photon dissociation for coronene and circumcoronene. This contribution is negligible for coronene but relevant for circumcoronene and causes $k$ to converge to a $d^{-2}$ proportionality at distances larger than 100\,AU.
We fit the numerical values with analytical multi power-laws and apply the fit values to the disk model.\\
\\Figure \ref{fig:k/Zn} shows a comparison between the PAH monomer collision time scale (\textit{filled contours}) and the PAH dimer dissociation time scale (\textit{black contours}, same units) for the three simulated monomers coronene (C$_{24}$H$_{12}$), circumcoronene (C$_{54}$H$_{18}$) and circumcircumcoronene (C$_{96}$H$_{24}$).
In all cases, the collision time scale for monomers is significantly shorter, $\approx$ seconds to yrs, than the typical lifetime of a disk of a few Myrs \citep{Haisch2001}.
There is no significant difference of monomer collision rates between the Herbig star disk and the T\,Tauri star disk.
In contrast, the dissociation time scale for dimers differs between the Herbig model and the T\,Tauri model.
Because of the higher luminosity and the more energetic photons, the dissociation time scale in the Herbig model is several orders of magnitude lower than for the T\,Tauri model.
While coronene and circumcoronene can be dissociated in less than a year within the innermost 100\,AU, circumcircumcoronene cannot be dissociated beyond 50\,AU in the Herbig model and 7\,AU in the T\,Tauri model.\\
\\Optical depth was not considered in deriving the cluster dissociation time scales. 
For further analysis, we indicate the estimated $\tau_\text{UV}=1$ surface for several effective absorption cross sections (\textit{red, blue, green lines}) to separate the disk into the photosphere and the sub-photospheric layer.
In this simple separation, the photosphere will be mostly dominated by dissociation events. Only for circumcircumcoronene in the Herbig model beyond 20\,AU, collisions occur at a higher rate than dissociation. For the T\,Tauri model, circumcoronene and circumcircumcoronene monomer collisions will occur faster than a dimer dissociates beyond 30\,AU resp. 2\,AU. The sub-photospheric layers are collision dominated and allow cluster to grow beyond the dimer stage because of the high collision rate.
We further want to analyse the stability of PAH clusters as a function of cluster size given that the collision time scales are rather short.

\subsection{PAH Cluster Size Effects}
\label{sec:clusterstability}
To investigate the stability of larger cluster, we perform our Monte-Carlo calculations for clusters with up to 200 monomers and track the dissociation rates for both disks and all three PAH molecules. Given that the heat capacity increases approximately linear with the number of carbon atoms in a cluster and each carbon atom absorbs the same amount of energy, cluster have the same radiative equilibrium temperature independent of size. Therefore, it is more practical to present temperatures instead of internal energies to explain the effects on the dissociation rate. Depending on the radiation field intensity, a growing cluster can either increase its stability if $T_\text{eq}<T_\text{crit}$ or contrary weaken it if $T_\text{eq}>T_\text{crit}$.\\
\\In the first case where $T_\text{eq}<T_\text{crit}$, single photon events become very unlikely for larger cluster and the dissociation rate is entirely dominated by the temperature probability distribution.
Figure \ref{fig:GE_gauss} shows the temperature probability distribution for growing circumcircumcoronene clusters in the same radiative environment where the radiative equilibrium temperature is smaller than the critical temperature.
While a tetramer undergoes large temperature fluctuations, larger cluster have narrower temperature distributions with the maximum located at the radiative equilibrium value of $T_\text{eq}= 1073$\,K corresponding to $E_\text{eq,4}= 71$\,eV, $E_\text{eq,16}= 286$\,eV, $E_\text{eq,64}= 1144$\,eV and $E_\text{eq,256}= 4578$\,eV for the clusters shown and approximated by the analytical formulas given in appendix \ref{sec:analyticalG}.

\begin{figure}
    \centering
    \includegraphics[width=1.0\linewidth]{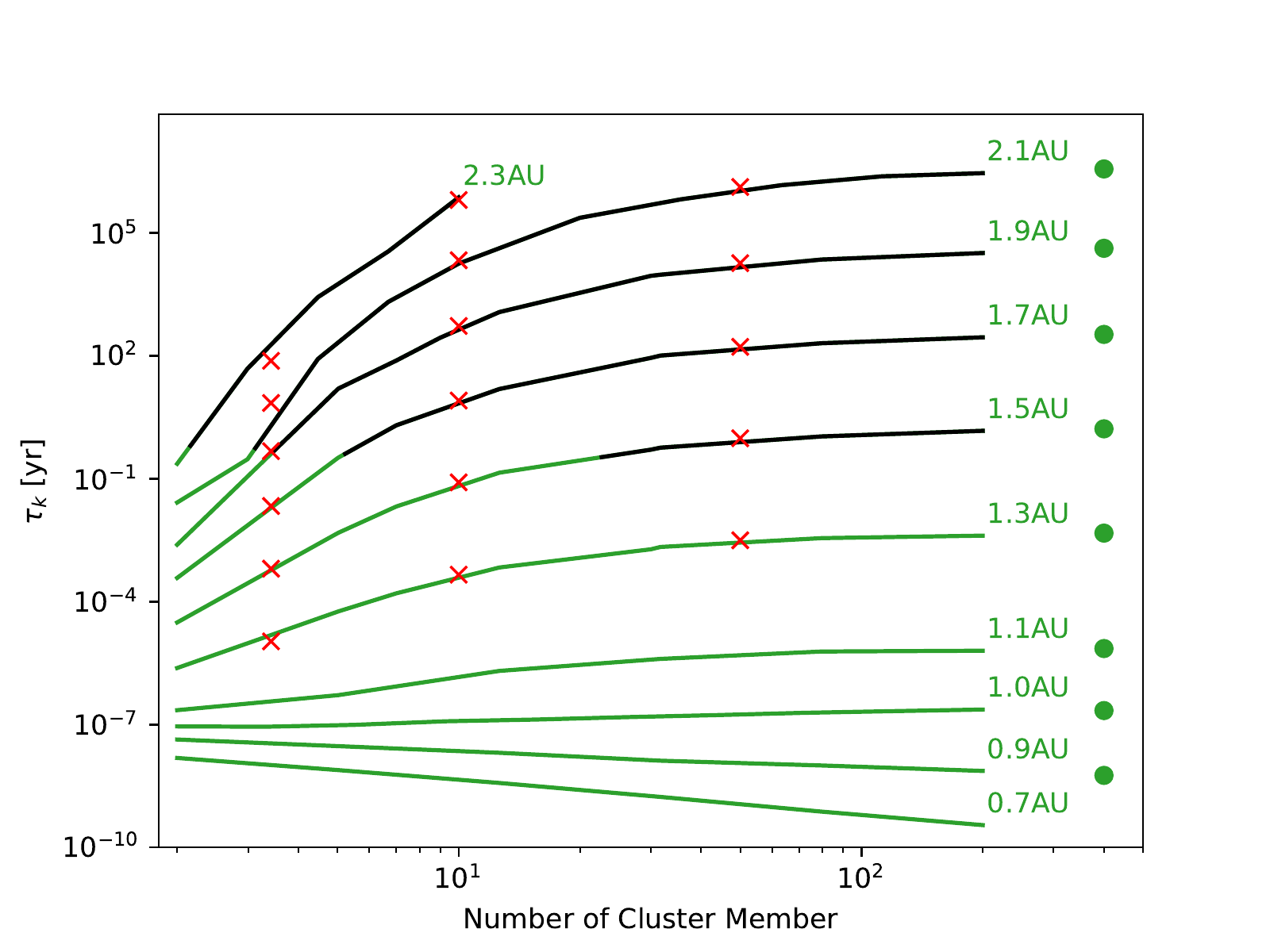}
    \caption{Dissociation time scale for circumcircumcoronene clusters in the T\,Tauri model. \textit{solid lines}: Monte-Carlo simulations \textit{crosses}: statistical calculations, \textit{Circles}: Value at radiative equilibrium temperature. The transition from green to black denotes the cluster size at which the vertical mixing time scale equals the dissociation time scale. Up to 0.9\,AU, the radiative equilibrium temperature is higher than the critical temperature for dissociation and cluster cannot gain stability. At larger distances, cluster stability increases with a few more cluster member.}
    \label{fig:CCCkNC}
\end{figure}

\begin{figure}
    \centering
    \includegraphics[width=1\linewidth]{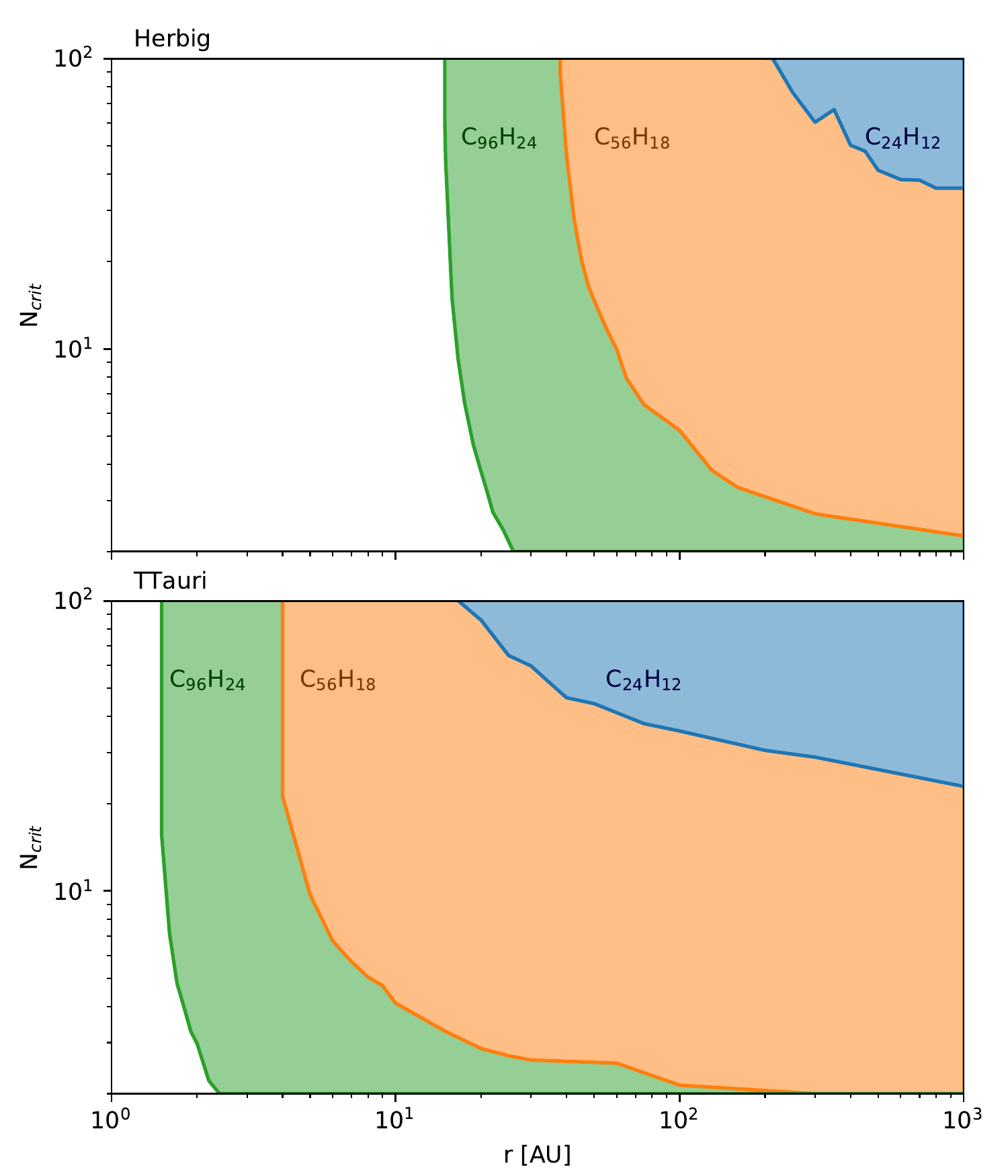}
    \caption{Critical number of cluster member where $t_\text{ed}=\tau_\text{k}$ for the Herbig model (\textit{top, lines}) and the T\,Tauri model (\textit{bottom, lines}). The smallest stable molecule is indicated by the fill colour. For the Herbig star model, cluster are stable outside the regions where they can possibly form. In the T\,Tauri disk model, circumcoronene and circumcircumcoronene can be stabilised where cluster can form in the midplane. Coronene cluster are unlikely to survive.}
    \label{fig:crit-cluster-member}
\end{figure}

The larger the cluster, the shorter the time interval between photon absorptions and the smaller the degree of temperature fluctuations.
Since high temperature events become less likely the larger the cluster is, the dissociation rate decreases and converges towards the radiative equilibrium value. In addition, the greatest increase in stability is gained by the smallest clusters.\\
\\Contrary, at smaller distances where $T_\text{eq}>T_\text{crit}$, cluster undergo dissociation before the equilibrium temperature can be reached.
After an initial dissociation event, the time until the next dissociation is determined by the time it takes to compensate for the energy loss of the previous dissociation. The smaller the cluster, the more energy is lost by release of a monomer relative to the total energy of the cluster (see eq. \eqref{eq:energy_fraction}). However, the absorbed energy flux is proportional to the number of molecules in the cluster and therefore smaller cluster need longer to compensate for the energy loss. Hence, cluster cannot be stabilised in very strong radiation fields and are even faster dissociated the larger they are.\\
\\\\Figure \ref{fig:CCCkNC} shows the dissociation time scale as a function of cluster size for circumcircumcoronene in the T\,Tauri model. 
Following the idea of \citet{Siebenmorgen2010}, we want to compare our dissociation rates to the vertical mixing time scale at a given distance.
We estimate the local eddy time scale $t_\text{ed}$ similar to \citet{Dullemond2004} with q = 0.5 by

\begin{equation}
    t_\text{ed} = \frac{\alpha^{1-2q}}{\Omega_\text{k}}
\end{equation}
where $\alpha$ is the alpha viscosity parameter first described by \citet{Shakura1973} and $\Omega_\text{k}$ is the local Keplerian orbital velocity.
Where mixing is faster than dissociation ($\tau_\text{k}>t_\text{ed}$), the coloured lines transition from green to black. We call these clusters stable implying that a fraction of the clusters is able to survive.
The dissociation rates for very large clusters are noted as circles. 
At small distances, cluster do not gain stability by increasing cluster size and undergo cascade dissociation events in a short period of time.
Although dissociation still occurs on short time scales at 1\,AU, clusters will gain stability with larger sizes because $T_\text{eq}$ declines below $T_\text{crit}$ so that the dissociation is entirely fluctuation dominated.
Then, the dissociation time scale will approach the value at equilibrium temperature for growing clusters. Outside 2\,AU, circumcircumcoronene dimers are stable. 
The full diagram with all simulated PAHs for both disk models can be found in appendix (figure \ref{fig:k(Nc)}).\\
\\To summarise our results, we track the cluster size where $t_\text{ed} = \tau_\text{k}(N_\text{crit})$ for all simulated PAHs and distances. Figure \ref{fig:crit-cluster-member} shows the critical cluster size as a function of distance when exposed to the full stellar flux. The fill colour represents the smallest PAH that is stable at a given distance. The transition from unstable to stable of dimers occurs over a short distance for circumcircumcoronene. Circumcoronene can be stabilised but needs to grow beyond dimers in most parts of the disks.
The smallest molecule coronene needs at least 30 cluster members to gain stability in the outer parts of the disk. 
For the Herbig model, no cluster is expected to survive in the innermost 20\,AU independent of size and PAH. In the T\,Tauri model, no cluster can survive within 1.5\,AU.

\section{Discussion}
\label{sec:discussion}

In our study we find that PAH cluster formation can occur within years in the entire disk if the cluster is shielded from stellar radiation. 
Studies report that PAH cluster gain additional vibrational lines \citep{Rapacioli2007} and may be associated with the broad non-continuum plateaus underlying the PAH feature emission \citep{Bregman1989, Rapacioli2005b}. 
PAH abundance measurements such as \citet{Acke2010} are based on extraction of sharp PAH features. 
Under the condition that PAH cluster are mixed from the midplane into the photosphere and survive, they could contribute to the underlying continuum and PAH feature flux studies might underestimate the total amount of PAH molecules in the protoplanetary disk. We therefore want to discuss this possibility in connection with our results.

\subsection{Herbig Model}
The comparison between the cluster formation and the stability of clusters (fig. \ref{fig:crit-cluster-member}) shows, that coronene and circumcoronene cluster will break-up within a short time when exposed to the full stellar radiation and do not survive. Circumcircumcoronene cluster are able to withstand the stellar radiation outside 25\,AU by forming dimers, circumcoronene quadrumers can withstand the stellar radiation beyond 100\,AU. These sizes can be reached within a couple of years. Hence, we expect cluster to be present in the photosphere at large distances.
As HD169142 is on the cooler and less luminous end of the Herbig stars, the dissociation rates can be treated as a lower limit for the Herbig stars.
We therefore conclude that PAHs are prominently present as monomers in the inner disk's photosphere. We cannot rule out that cluster formation can explain that no PAH emission is detected in 40\,\% of Herbig star disks but consider this scenario unlikely.
An alternative explanation for the missing PAH flux is likely the high UV flux that causes PAH monomers to be irreversibly destroyed and deplete PAHs in disks.

\subsection{T\,Tauri Model}
In contrast to the Herbig model, PAH cluster can be stabilised at small radii in the T\,Tauri disk if they have the size of circumcoronene. Based on the monomer collision rate, cluster can grow to this size within a couple of years. These cluster would be stable to the stellar radiation and could grow further even if exposed to stellar radiation. Hence, the formation of PAH cluster can be one possible explanation why only 10\,\% of T\,Tauri disks show PAH features. In the literature, attenuation and lack of excitation emission have been excluded while cluster formation and trapping of PAHs in ices have been raised as possible explanations by \citet{Geers2009}. Cluster formation and desorption on dust grains are competing in the sub-photospheric layers. To trace the importance of both processes, a PAH time evolution model is required that takes collisions between PAHs and dust grains into account. Furthermore with a time evolution model, cluster growth could be tracked and linked to the strength of vertical mixing. We will consider these processes in a future study.\\
\\Furthermore, our derived dissociation rates can be extrapolated to desorption of PAHs from dust grains. If we consider dust grains as very large cluster at radiative equilibrium temperature, then the rate at which PAH monomers detach from the surface of a grain is given by the equilibrium rate indicated by circles in figures \ref{fig:CCCkNC} and \ref{fig:k(Nc)}. Therefore, if a PAH cluster of any size is stable, individual PAHs will also stay adsorbed on dust grain surfaces. This approximation might be also valid for small PAH clusters but for larger sizes PAH clusters will likely behave like small grains rather than individual molecules. Interpreting figure \ref{fig:crit-cluster-member}, adsorption of PAHs is rather important for T\,Tauri star disks than for Herbig star disks.\\
\\In our study, we have a made multiple approximations and here we will discuss their effect on our results. Rather than calculating the emission probabilities for each PAH band at all temperatures, we decided to use a simplified approach by using directly the cooling law obtained by the emission model of \citet{Bakes2001} and a constant IR photon energy. Since we calculate the emission rate directly from the cooling law, the choice of the emission photon energy is arbitrary but sets the resolution and noise level of the resulting energy distribution. Higher photon energies reduce the noise level of $G(E)$ at the cost of lower resolution while lower photon energies have the reversed effect. We find that the choice $E_{IR}$ in the range of physical PAH bands has only a minor effect on the computed dissociation rate which is mostly dominated by the Monte-Carlo noise at the very rare high energies. \\
\\Furthermore, we have assumed in section \ref{sec:N=N+1} that a cluster dissociating from size $N_\text{mem}$ will have a similar energy compared to a cluster of size $N_\text{mem}+1$ after dissociation. This approximation holds well for large cluster which are close to radiative equilibrium because their absolute energy loss due to dissociation becomes almost constant. Our approximation gets worse for smaller cluster because the relative energy loss is proportional to $N_\text{mem}/(N_\text{mem}+1)$. Hence, we are overestimating the relative energy loss from dissociation by $(N_\text{mem}^2 + 1)/N_\text{mem}^2$ and underestimating the dissociation rate. For a dimer formed from dissociating a trimer, the relative energy difference is 25\%. The resulting change in the dissociation rate is difficult to estimate because of the exponential behaviour of the Arrhenius law and because the previous dissociation event is only a fraction of the total energy probability distribution. Nevertheless, this effect is only strong when the PAHs undergo quick cascading dissociation. In these cases, the dissociation rate is very high and dominates over clustering. Therefore, uncertainties in the dissociation time scale do not affect the long-term evolution of PAH clusters.\\
\\In equation \eqref{eqEA} we have to assume the binding energy of a monomer to a cluster that is used for all cluster sizes. As stated before, because of the exponential dependence on $E_\text{A}$, small changes in $E_\text{A}$ can have large impacts on the dissociation rate at a given energy. We find that a deviation of $\pm$10\,\% in $E_\text{A}$ can change the local dissociation rate by a factor two for very high dissociation rates ($k\approx 1\,\text{s}^{-1}$) to two orders of magnitude when the dissociation rate is low ($k\approx 10^{-13}\,\text{s}^{-1}$) depending on the exact simulation configuration including cluster size, PAH species and radiation field. Although this is locally a huge effect, the steep gradient of $k(r)$ compensates this rapidly. An decrease of the dissociation rate by two orders of magnitudes in figures \ref{fig:CCCkNC} and \ref{fig:k(Nc)} corresponds to a maximum increase in distance of 50\,AU for large coronene clusters in the Herbig model and a minimum increase of 0.2\,AU for large circumcircumcoronene clusters in the T\,Tauri model. Compared to the individual distances where cluster stability is achieved in our conclusion figure \ref{fig:crit-cluster-member}, this corresponds to a maximum shift for coronene in the Herbig model from 200\,AU to 250\,AU and a minimum shift from 1.5 to 1.7\,AU for circumcircumcoronene in the T\,Tauri model. \\
\\For a full time-evolution model we want to discuss further processes that are relevant to include but have not been considered in this work. As reported by \citet{Siebenmorgen2010}, the X-ray excess of T\,Tauri stars is an additional energy source and can penetrate deeper into the disk compared to UV radiation \citep{Ryter1996}. The absorption of X-rays by PAHs has been studied in the laboratory e.g. by \citet{Reitsma2014}. Furthermore, their secondary effects as well as secondary effects of cosmic rays can produce additional UV photons locally in the sub-photospheric layers \citep{Tielens2005} that can potentially destroy small PAH cluster via single photon dissociation.\\
\\In our analysis of the photo fragmentation, we have assumed that IR emission is the only stabilizing process. However, for large PAHs, the monomer binding energy approaches that of covalent bonds of e.g., C-H. To compare our derived rates for dimers (figure \ref{fig:k(E)}) to the dehydrogenation rates for monomers obtained by \citet{Andrews2016}, we need to extrapolate their internal energies by multiplication with a factor 2 to correct for the additional degrees of freedom of the dimer. While coronene and circumcoronene dimers will dissociate rather than release H or H$_2$, circumcircumcoronene will break H-C bonds before a dimer will dissociate. For high atomic H densities, e.g. in the upper photosphere,  dehydrogenation-hydrogenation cycles would act to stabilize clusters made of large PAHs.\\
\\Lastly, ionisation has not been considered yet. Typical ionisation potentials for neutral PAH monomers are on the order of 5-7\,eV \citep{Tielens2005} and \citet{Joblin2017} report decreasing ionisation potentials for growing PAH cluster. \citet{Rapacioli2009} report that the activation energy for dissociation of small, ionised cluster is significantly higher than of neutral cluster. Therefore, ionisation can support the stability of PAH cluster by competing with dissociation while increasing the energy needed to dissociate ionised clusters especially for small clusters that are dominantly dissociated by single photon events. As a side effect, ionisation decreases the probability of reaching higher energies in PAHs that experience large energy fluctuations reducing the strength of short wavelength emission. This requires sufficient free electrons to recombine in order to keep the neutral fraction high and is probably limited to small regions in a protoplanetary disk. Whether these processes support cluster survivability in the photosphere of Herbig and T\,Tauri stars needs to be investigated in a future study.


\section{Conclusions}
\label{sec:conclusions}
We have presented calculations for cluster formation of PAHs under protoplanetary disk conditions for a Herbig Ae/Be disk model and a T\,Tauri disk model assuming a pristine PAH abundance similar to the ISM. We report that dimer formation happens on time scales much shorter than a disk life time for molecules of astrophysical relevance in the sub-photospheric layers of protoplanetary disks. Cluster are unlikely to survive in the inner disk photosphere and need to be carried to the photosphere by vertical mixing or other processes. We further provide calculations for dissociation rates of different cluster sizes depending on the radiation field intensity. For the Herbig star disk model, cluster are unlikely to survive because of the strong stellar radiation if no other energy loss channel than IR emission is considered. Under these conditions, cluster formation is unlikely to be the reason why 40\,\% of Herbig stars do not show PAH emission. For the T\,Tauri disk model, PAHs as large as circumcoronene or larger are able to form cluster which can withstand the stellar UV radiation. We propose PAH cluster formation as one possible explanation for the lack of PAH emission in most T\,Tauri disks. For further investigation, a more sophisticated PAH evolution model has to be developed taking into account competing processes to dissociation such as ionisation and dehydrogenation.\\

\begin{acknowledgements}
The authors thank the anonymous referee for suggestions to improve the quality of the paper. K.L. acknowledges funding from the Nederlandse Onderzoekschool Voor Astronomie (NOVA) project number R.2320.0130. C.D. acknowledges funding from the Netherlands Organisation for Scientific Research (NWO) TOP-1 grant as part of the research program “Herbig Ae/Be stars, Rosetta stones for understanding the formation of planetary systems”, project number 614.001.552. 
Studies of interstellar PAHs at Leiden Observatory are supported through a Spinoza award from the Dutch research council, NWO.
\end{acknowledgements}

\bibliographystyle{aa} 
\bibliography{library.bib} 
\appendix
\section{Derivation of $k_0$}
The derivation of the pre-exponential factor $k_0$ for the Arrhenius law for dissociation of single molecule from a cluster is based on \citet{Tielens2005} and \citet{Tielens2020}. From thermodynamics, the rate constant can be expressed with
\begin{equation}
    k_0 = \frac{k_\text{B}T}{h} \frac{Z_\text{vib}^\text{f}Z_\text{rot}}{Z_\text{vib}^\text{i}}
\end{equation}
assuming a monoatomic gas. $k_\text{b}$ describes Boltzmann's constant, $T$ is the temperature, $h$ is Planck's constant and $Z$ describes the vibrational and rotational partition functions. We assume that only one molecule can detach from the cluster at a time and the number of degrees of freedom by the breakup of the cluster stays constant. If we consider only the perpendicular mode for excitation, we can write $k_0$ as
\begin{equation}
    k_0 = \sqrt{\pi}\nu_\text{vdw}\left(\frac{k_\text{B}T}{hcB}\right)^{3/2}
\end{equation}
where $\nu_\text{vdw}$ is the frequency of the van der Waals bond and $B$ is the rotational constant. Knowing that $\Delta J = +1$ and $\Delta J = -1$ transitions during an de-excitation cascade are balanced leads to 
\begin{equation}
    2J = J_\text{IR} = \left(\frac{hc\bar{\nu}}{6hcB}\right)^{1/2}
\end{equation}
with $\bar{\nu}$ being the typical IR photon energy.
Together with the rotational excitation temperature
\begin{equation}
   T_\text{rot} = \frac{hcBJ_\text{IR}^2}{ k_\text{B}}
\end{equation}
which finally leads to
\begin{equation}
    k_0 = \sqrt{\pi} \nu_\text{vdw} \left(\frac{\bar{\nu}}{24B}\right)^{3/2}\,.
\end{equation}
Using typical values for PAHs of $\nu_\text{vdw}$ = 50 - 70\,cm$^{-1}$ results in $k_0 = 2.5 \cdot 10^{17}$ s$^{-1}$.

\section{Analytical Calculation of $G(E)$}
\label{sec:analyticalG}
When the photodissociation rate is low and the photon absorption rate is sufficiently high ($\tau_{UV} \not\gg \tau_\text{cool}$), the energy probability distribution can be approximated by a Gaussian distribution \citep{Tielens2020} that is a result of the Poissonian character of photon absorption. The accuracy of this approach increases the closer the cluster to the radiative equilibrium is and the smaller the energy fluctuations. Then, the mean of the distribution is equal to the radiative equilibrium energy $E_\text{eq}$ given by
\begin{equation}
    E_\text{eq}: \frac{\text{d}E_\text{eq}}{\text{d}t} = \overline{E_\text{UV}}\,\, \overline{k_\text{UV}}
\end{equation}
where $\overline{E_\text{UV}}$ is the mean UV photon energy and  $\overline{k_\text{UV}}$ is the equivalent mean photon absorption rate. The Gaussian standard deviation $\sigma$ can be calculated with
\begin{equation}
    \sigma_E = \kappa \sqrt{E_\text{eq}\overline{E_\text{UV}}}.
\end{equation}
Note that $\kappa$ is an empirical correction factor necessary to account for the non-Poissonian characteristic of the cooling law. The correction factor is dependent on the exact shape of the cooling law. We found a suitable value of $\kappa\approx6/11$ for our simulated cases. Unfortunately, this relation is not suitable to evaluate eq. \eqref{eq:k_int} for small cluster because of the asymmetrical character of $G(E)$ and the exponential dependence of $k$ on the energy.
\section{Fitting parameters for $k(d)$}
All data points in figure \ref{fig:k(d)} are fitted in logarithmic space with linear functions of shape
\begin{equation}
    \text{log}_{10}(k) = k_i + m_i \text{log}_{10}(d)
\end{equation}
which corresponds to 
\begin{equation}
    k = 10^{k_i} \cdot d^{m_i}
\end{equation}
in linear space using $k$ in s and $d$ in AU. The individual power laws are connected similar to equation \eqref{eq:T(E)} where the power has to be adjusted by hand to match the correct shape.

\begin{table}[h]
    \caption{Parameters obtained from fitting dissociation rates of dimers in figure \ref{fig:k(d)} as a function of distance. Where a $d^{-2}$ can be physically explained, we exclude $m_i$ as a fitting parameter.}
    \begin{tabular}{l|l|r|r|r|r|r|r}
         Model & Monomer & $k_1$ & $m_1$ & $k_2$ & $m_2$ & $k_3$ & $m_3$\\
         \hline
         Herbig & C$_{24}$H$_{12}$ & 2.4 & -2 & & & & \\
          & C$_{54}$H$_{18}$ & 2.3 & -2 & 8.4 & -6.8 & & \\
          & C$_{96}$H$_{24}$ & 2.1 & -2 & 26.2 & -24.5 & & \\
         T\,Tauri & C$_{24}$H$_{12}$ & 0.2 & -2 & & & & \\
          & C$_{54}$H$_{18}$ & 0.4 & -2 & 1 & -5.4 & -5.7 & -2 \\
          & C$_{96}$H$_{24}$ & 0.2 & -2 & 0.4 & -19.7& & \\
    \end{tabular}

    \label{tab:my_label}
\end{table}
\clearpage

\onecolumn
\begin{sidewaysfigure}[ht]
    \includegraphics[width=1\linewidth]{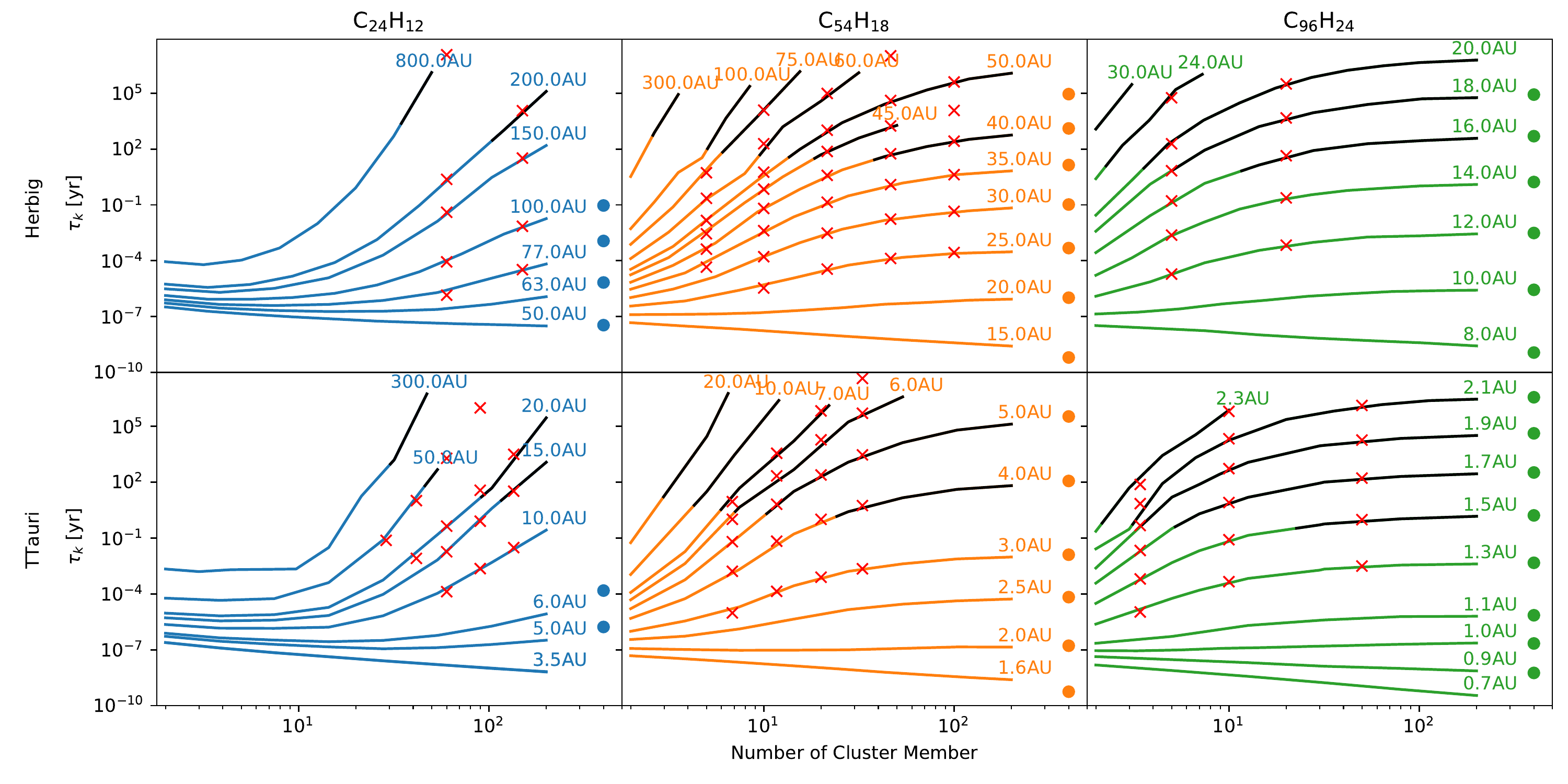}
    \caption{Full plot for figure \ref{fig:CCCkNC}. \textit{Top row}: Herbig model; \textit{Bottom row}: T\,Tauri model. \textit{From left to right}: coronene, circumcoronene and circumcircumcoronene. \textit{solid lines}: values obtained from MC-simulations, \textit{crosses}: values obtained from statistical calculations. The transition from coloured lines to black shows the cluster size at which the dissociation time scale of the cluster equals the eddy turnover time scale. We consider black lines as stable. The smaller the monomer molecule, the larger the distances required to reach stability and the smaller the stability gain per member molecule. Coronene is too small to be stabilised over 1000\,AU in both disks.}
    \label{fig:k(Nc)}
\end{sidewaysfigure}
\end{document}